\documentclass[12pt]{article} \topmargin -5mm \oddsidemargin 7mm
\usepackage{graphicx,multirow}
\usepackage{amssymb,amsmath,amsfonts,bm}
\usepackage{colortbl}
\usepackage{natbib}
\usepackage[flushleft]{threeparttable}
\usepackage{tablefootnote}

\title{\bf Multi-Parameter Regression Survival Modelling with Random Effects}
\author{Fatima-Zahra Jaouimaa\\\emph{University of Limerick, Ireland}\\
	Il Do Ha\\\emph{Pukyong National University, South Korea}\\
Kevin Burke\\\emph{University of Limerick, Ireland}\\}

\date{}

\begin{document}
\maketitle
\bigskip

\begin{abstract}
We consider a parametric modelling approach for survival data where covariates are allowed to enter the model through multiple distributional parameters, i.e., scale and shape. This is in contrast with the standard convention of having a single covariate-dependent parameter, typically the scale. Taking what is referred to as a multi-parameter regression (MPR) approach to modelling has been shown to produce flexible and robust models with relatively low model complexity cost. However, it is very common to have clustered data arising from survival analysis studies, and this is something that is under developed in the MPR context. The purpose of this article is to extend MPR models to handle multivariate survival data by introducing random effects in both the scale and the shape regression components. We consider a variety of possible dependence structures for these random effects (independent, shared, and correlated), and estimation proceeds using a h-likelihood approach. The performance of our estimation procedure is investigated by a way of an extensive simulation study, and the merits of our modelling approach are illustrated through applications to two real data examples, a lung cancer dataset and a bladder cancer dataset.
\end{abstract}

\section{Introduction\label{sec:intro}}

Standard survival analysis models such as the proportional hazards (PH) model are known to have a single covariate-dependent parameter, the scale parameter. As a means to extend these models and afford more flexibility in the modelling of survival data, \citet{burke2017multi} developed a multi-parameter regression (MPR) approach for a parametric hazard function. MPR is an approach whereby more than one distributional parameter is allowed to depend on covariates, and this is sometimes referred to as ``distributional regression'' \citep{stasinopoulos2018gamlss}; see also \citet{rigby2005generalized} and \citet{stasinopoulos2007generalized}. Using the Weibull MPR model as an example, Burke and Mackenzie's paper demonstrates the advantages of allowing both the scale and the shape parameters to depend on covariates simultaneously. More recently, \citet{burke2020semiparametric} extended a semi-parametric accelerated failure time (AFT) model to multiparameter regression and \citet{burke2020flexible} explored an MPR parametric survival modelling framework in which the baseline hazard function follows an adapted power generalized Weibull (APGW) distribution. 
\par
Standard MPR models rely on the assumption of independent event times, and such an assumption cannot be met in studies with clustered observations. This includes studies where successive or recurrent event times are recorded on each subject; multi-centre studies where survival times of individuals from the same centre may be dependent due to centre-specific conditions, clinical or otherwise; family studies; and matched pair studies. Various methods have been developed to model the lack of independence in clustered data, but perhaps the most standard approach is the one whereby a cluster-specific random effect is introduced into the model. Given these random effects, the data are assumed to be conditionally independent. In the context of survival analysis, such random effects models are commonly referred to as frailty models and since the random effect represents a common effect on all members in a cluster, these types of common risk models are known as shared frailty models \citep{clayton1978model, duchateau2007frailty, hougaard2012analysis}. 

Extensions of the MPR models to include random effects have been centred around the ``classical'' multiplicative frailty model, whereby the frailty term is included in the linear predictor corresponding to the scale parameter. \citet{peng2020multiparameter} extend a Weibull MPR model for interval censored data to include a gamma distributed multiplicative frailty term, but only for univariate frailty. In their paper, \citet{jones2020bivariate} extend the MPR PGW and MPR APGW models to handle bivariate data using multiplicative frailty, where the dependence is understood using links to the well known power variance copula \citep{goethals2008frailty, duchateau2007frailty, hougaard2012analysis}. The attraction of MPR modelling is the flexibility afforded by allowing the hazard shapes to differ, but just as the shape can vary according to covariates, so too could it have random variation, for example, resulting from differences across multiple centres. Hence, in this paper, we go beyond the classical multiplicative model to allow for a wider range of model structures, specifically, models in which a frailty term is included in each distributional parameter. This, we believe, is a more natural way of modelling correlated data using MPR models than multiplicative frailty which relates only to the scale of the hazard. Furthermore, our proposed model accounts for the potential correlation between the frailty terms themselves since scale and shape effects may be correlated. 

Parameter estimation in the parametric shared frailty model is commonly achieved through integrating out the frailties from the conditional survival likelihood. The resulting equation is an explicit expression for the marginal likelihood, containing the fixed parameters and no longer the frailties. This marginal likelihood can then be maximized using numerical methods to yield estimates for the fixed effects. Inference about the random effects is not readily available and integrating out the frailties from the joint density typically involves the evaluation of analytically intractable integrals over the random-effect distributions. Numerical integration methods such as the Gauss-Hermite (G-H) quadrature can be used to approximate the value of the integrals, however, when the dimensionality of the integral is high, the number of quadrature points grows exponentially with the number of random effects, and the approximation is sub-optimal. While several methods, such as the Monte Carlo Expectation-Maximisation, Markov Chain Monte Carlo and Gibbs sampling have been used to overcome the issue of intractable integrals, these methods are notoriously known for being computationally intensive. This is especially true when the number of random effects (clusters) is large or when the complex correlation structure among the clustered survival times requires the assumption of multiple frailties \citep{vaida2000proportional, ripatti2000estimation, abrahantes2007comparison, duchateau2007frailty}. We adopt a so-called ``hierarchical likelihood'' (h-likelihood) approach. This approach was originally proposed by \citet{lee1996hierarchical} for a generalized mixed-effects model but further studied and developed to provide a straightforward, unified framework for various random effect models including frailty models \citep{ha2001hierarchical, do2002hierarchical, ha2003estimating, do2017statistical}. 

In contrast to the standard marginal likelihood approach, the h-likelihood framework treats the random effects or frailties as model parameters, which are then jointly estimated with the fixed parameters and the frailty dispersion parameter(s). Estimates of the fixed and random effects are found by maximising the log-likelihood function conditional on the random effects plus a penalty term whose value depends on how dispersed the random parameters are, i.e., if the random effects have a large dispersion parameter, then the penalty term takes a large value. By treating the random effects as model parameters, the h-likelihood framework avoids the intractable integration needed to calculate the marginal likelihood and provides an efficient estimation procedure. Furthermore, classical analysis of random effects models focuses on the estimation of the fixed parameters and the frailty variance parameter(s), but in many recent applications, estimation of the random effects is also of interest. Such estimates allow for the survivor function for individuals with given characteristics to be estimated, i.e., the cluster specific failure time distribution, and are especially useful in multi-centre studies when the frailty term represents the centres. Estimates of the random effects in such studies provide information about the merits of the different centres in terms of patient survival and are useful for investigating the potential heterogeneity in survival among clusters in order to better understand and interpret the variability in the data \citep{ha2016interval}. 

The remainder of this article is organized as follows: Section 2 describes the MPR model, its extended version with random effects and the h-likelihood procedure for parameter estimation in the given model. Section 3 presents results of extensive simulation studies. The proposed methods are illustrated on two datasets arising from multi-centre studies and the results are shown in Section 4, followed by a discussion and conclusion in Section 5.
\section{The MPR Frailty model} \label{sec:MPR_frailty_mod}
\subsection{Model Formulation}
To formulate a shared frailty MPR model for survival data, we assume time-to-event data arising from a multi-centre study. In this study, we have $q$ centres or clusters, with $n_i$ individuals (patients), $i = 1, 2, 3,\dots, q$. The total sample size is the total number of individuals coming from all $q$ centres, i.e., $n = \sum_{i=1}^{q}n_i$. We define $\tilde{T}_{ij}$ as the survival time for the $j$th individual, $j = 1, 2, 3,\dots, n_i$, in the $i$th cluster and $C_{ij}$ as the corresponding censored time. The cumulative hazard function for a shared frailty MPR model takes the following parameteric form:
\begin{equation} \label{eq:MPR-BVN}
	\Lambda(t_{ij}|x_{ij}, v_{\beta i}, v_{\alpha i}) = \tau_{ij}\Lambda_0(t_{ij}^{\gamma_{ij}}),
\end{equation}
where $\Lambda_0(.)$ is the underlying cumulative hazard function with scale parameters 
$\tau_{ij} > 0$ and shape parameter $\gamma_{ij} > 0$. The corresponding hazard function is given by $$\lambda(t_{ij}|x_{ij}, v_{\beta i}, v_{\alpha i}) =\tau_{ij}\gamma_{ij}t_{ij}^{\gamma_{ij}-1}\lambda_0(t_{ij}^{\gamma_{ij}}),$$ where $\lambda_0(.)$ is the baseline hazard function. $\Lambda_0(.;\gamma_{ij})$ can be one of the commonly used survival distributions (Weibull, Gomperts, or log-logistic), see Table \ref{tab:dists}. 

\begin{table}[h]
	\caption{Possible distributions.}\label{tab:dists}
	\begin{center}
		\begin{tabular}{lc}
			\hline
			$\Lambda_0(t)$ & distribution \\
			\hline
			$t$ & Weibull \\
			$\exp(t) - 1 $& Gompertz\\
			$\log(1 + t)$ & log-logistic\\
			\hline
		\end{tabular}
	\end{center}
\end{table}

The two distributional parameters, $\tau_{ij}$ and $\gamma_{ij}$ depend on covariates as follows: $$\log(\tau_{ij}) = {x}_{ij}^T{\beta} + v_{\beta i}, \qquad \log(\gamma_{ij})={x}_{ij}^T{\alpha} + v_{\alpha i},$$
where ${x}_{ij}=(1,x_{\beta ij1},\dots,x_{\beta ijp})^T$ is the covariate vector, ${\beta}=(\beta_0,\beta_1,\dots,\beta_p)^T$ and ${\alpha}=(\alpha_0,\alpha_1,\dots,\alpha_p)^T$ are the corresponding regression coefficient vectors, $v_{\beta i}$ and $v_{\alpha i}$ denote the scale and shape random effects from the $i$th cluster respectively, and to ensure positivity of the parameters a log link is used. Although we allow the scale and the shape parameters to depend on the same set of covariates, the parameters may or may not have covariates in common depending on the value of the corresponding regression coefficients. The individual cluster-specific effects (scale or shape random effects) are assumed to be independently and identically distributed (i.i.d.) according to some distribution. 

In their work, \citet{burke2017multi} found the distributional parameters in an MPR model to be highly correlated. To account for the possibility of the propagation of this correlation to the corresponding frailty terms, we allow for correlation between the two random effects terms, $v_{\beta i}$ and $v_{\alpha i}$, by assuming that ${v}_i = (v_{\beta i}, v_{\alpha i})^T$ follow a bivariate normal distribution such that 
$$ {v}_i = \Bigg( \begin{matrix} v_{\beta i}\\ v_{\alpha i} \end{matrix} \Bigg) \sim BVN \Bigg(\Bigg(\begin{matrix}
	0\\
	0
\end{matrix}\Bigg)
,\ {\Sigma} = \Bigg(\begin{matrix}
	\sigma_{\beta}^2 & \rho \sigma_\beta \sigma_\alpha\\
	\rho \sigma_\alpha \sigma_\beta & \sigma_{\alpha}^2 
\end{matrix}\Bigg) \Bigg).$$

Fitting the model which assumes the bivariate normal distribution for the frailty terms avoids the assumption of independence which may be unnecessarily restrictive. Furthermore, this model generalizes various models and simplifications of it by assuming null components or a specific correlation structure include: 

\begin{itemize}
	\item {${\alpha = 0, \sigma^2_{\beta} = 0, \sigma^2_{\alpha} = 0}:$} the proportional hazards (PH) model \citep{cox1972regression};
	\item {${\alpha = 0, \sigma^2_{\alpha} = 0}:$} a multiplicative frailty PH model \citep{duchateau2007frailty};
	\item {${\sigma^2_{\beta} = 0, \sigma^2_{\alpha} = 0}:$} a standard MPR model \citep{burke2017multi, burke2020flexible};
	\item {${\sigma^2_{\alpha} = 0}:$} a multiplicative scale frailty MPR model \citep{peng2020multiparameter, jones2020bivariate};
	\item {${\sigma^2_{\beta} = 0}:$} a shape frailty MPR model;
	\item {${\rho = 0}$}: we assume that the two frailty terms are independent and fit a model whereby $v_{\beta i}\sim N(0, \sigma_{\beta}^2)$ and $v_{\alpha i} \sim N(0,\sigma_{\alpha}^2)$; 
	\item {${\rho = \pm1}$}: $v_{\beta i}\sim N(0,\sigma_{\beta}^{2})$ and $v_{\alpha i} = \phi v_{\beta i}$ , where $\phi$ is a real-valued scaling factor \citep{do2017statistical}.
\end{itemize}

\begin{figure}[ht]
	\begin{center}
		\includegraphics[scale=0.45]{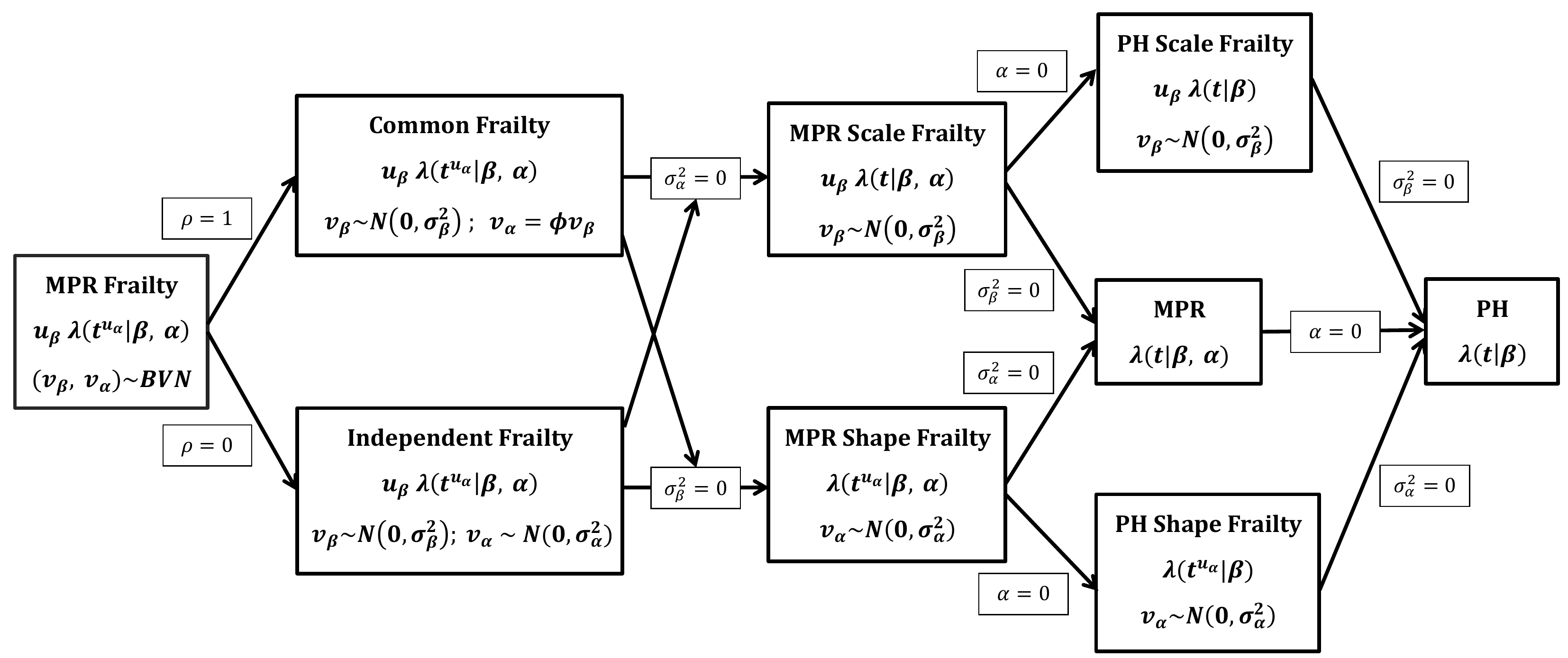}
	\end{center}
	\caption{A schematic diagram of some of the possible models generalized by the MPR frailty model assuming a bivariate normal distribution. Note that $u = \exp(v)$.} \label{fig:model_summary}
\end{figure}

Figure \ref{fig:model_summary} provides a summary of such models. Although, we only consider the normal distribution for the random effects, estimates of the fixed effects $({\beta}, {\alpha})$ are usually robust against violations of this assumption if the censoring rate or frailty variance parameter is not too high \citep{ha2001hierarchical, ha2003estimating, ha2011frailty, ha2016interval}.
\subsection{Construction of the H-Likelihood}
Denoting the observed data by the pairs ($T_{ij}, \delta_{ij}$), where $T_{ij} = \min(\tilde{T}_{ij}, C_{ij})$, the observed survival time for the $j$th individual in the $i$th cluster; and $\delta_{ij}$ is the censoring indicator, which takes the value 0 for a censored observation and 1 for an event. Under the standard assumptions that, given $v_\beta$ and $v_\alpha$, the censored times ($C_{ij}$s) and the event times ($\tilde{T}_{ij}$s) are conditionally independent and the censoring is conditionally non-informative, the log-h-likelihood function of the proposed model (\ref{eq:MPR-BVN}) is given by
\begin{equation}\label{eq:Hlike}
	\textit{h}= \textit{h}({\theta}, {\Sigma}) = \sum_{ij}\ell_{1ij} + \sum_{i}\ell_{2i},
\end{equation}
where
\begin{equation*}
	\ell_{1ij} = \ell_{1ij}({\theta}; t_{ij}, \delta_{ij}|v_i) = \delta_{ij}\{\log{\tau_{ij}} + \log{\gamma_{ij}} + (\gamma_{ij}-1) \log {t_{ij}} + \log{(\lambda_0(t_{ij}^{\gamma_{ij}}))}\} - \tau_{ij} \Lambda_0(t_{ij}^{\gamma_{ij}})
\end{equation*}is the logarithm of the conditional joint density function for $t_{ij}$ and $\delta_{ij}$ given the random effects $v_i = (v_{\beta i},v_{\alpha i})$, where $t_{ij}$ is the realisation of $T_{ij}$ and
\begin{equation*}
	\ell_{2ij} = \ell_{2ij}({\Sigma}; v_i) =  -\log{(2\pi \sigma_{\beta}\sigma_{\alpha} \sqrt{1-\rho^2})} - \frac{1}{2(1-\rho^2)} \bigg( \frac{v_{\beta i}^2}{\sigma_{\beta}^2}+\frac{v_{\alpha i}^2}{\sigma_{\alpha}^2} - 2\rho \frac{v_{\beta i} v_{\alpha i}}{\sigma_\beta \sigma_\alpha} \bigg) 
\end{equation*}
is the logarithm of the density function of $v_i$ with dispersion parameters (i.e. frailty parameters) ${\Sigma} = (\sigma_\beta, \sigma_\alpha, \rho)^{T}$, and ${\theta} = ({\beta}^T, {\alpha}^T)^T$ is the vector of the fixed parameters. For meaningful inferences, it is important to define the h-likelihood on a particular scale of ${v}$, such that the random effects occur linearly in the linear predictor. Hence, the log link we see in (\ref{eq:MPR-BVN}). \citep{lee1996hierarchical, ha2001hierarchical, lee2017generalized, do2017statistical}.
\subsection{Estimation Procedure} \label{sec:estimation}
From the h-likelihood function given in (\ref{eq:Hlike}), we can derive two likelihoods, namely the marginal likelihood and the restricted likelihood. The marginal likelihood eliminates the random effects, ${v}$, from $\textit{h}$ while the restricted likelihood eliminates the fixed effects ${\theta}$ from the marginal likelihood, i.e., eliminates both the fixed effects ${\theta}$ and the random effects ${v}$ from $\textit{h}$. In theory, the h-likelihood should be used for inference about ${v}$, the marginal likelihood should be used for inference about ${\theta}$ and the restricted likelihood for inference on the dispersion parameter(s) \citep{patterson1971recovery, harville1977maximum}. However, when the marginal likelihood is intractable, \citet{lee1996hierarchical, lee2001hierarchical} propose using adjusted profile likelihoods, $\textit{p}_{{v}}(\textit{h})$ and $\textit{p}_{{\theta}, {v}}(\textit{h})$ as approximations to the marginal likelihood and the restricted likelihood respectively. Consider the log-h-likelihood $\textit{h}$ with nuisance parameters ${\omega}$, the adjusted profile likelihood suggested by Lee and Nelder is given by
\begin{equation*}
	\textit{p}_{{\omega}}(\textit{h}) = \left[\textit{h} - \frac{1}{2}\log \det\{H(\textit{h};{\omega})/(2\pi)\} \right] \bigg|_{{\omega} = \hat{{\omega}}},
\end{equation*}
where $H(\textit{h};{\omega}) = -\partial^2 \textit{h}/\partial{\omega}^2$ and $\hat{{\omega}}$ solves $\partial \textit{h}/\partial{\omega} = 0$. The function $\textit{p}_{{\omega}}(.)$ produces an adjusted profile likelihood, profiling out the nuisance parameters ${\omega}$. The nuisance parameters can be the fixed effects and/or the random effects. $\textit{p}_{{v}}(\textit{h})$ is the first-order Laplace approximation to the marginal likelihood, $m$ \citep{lee2001hierarchical}. Similarly, $\textit{p}_{{\theta}}(m)$ is the first-order Laplace approximation to the restricted likelihood. Moreover, $\textit{p}_{{\theta}, {v}}(\textit{h})$ approximates $\textit{p}_{{\theta}}(\textit{p}_{{v}}(\textit{h}))$ and therefore, $\textit{p}_{{\theta}}(m)$. With the exception of binary data with a small cluster size (i.e., $n_i = 2$), it has been found that $\textit{h}$ and $\textit{p}_{{v}}(\textit{h})$ give very similar results when used in the estimation of the fixed effects \citep{lee2017generalized}. Hence, \citet{lee2017generalized} recommend the joint maximisation of $h$ over the fixed and random effects and refer to this as the uncorrected h-likelihood method. This methods works well for various models \citep{do2002hierarchical, do2017statistical}, including the models we propose.

The h-likelihood function given in (\ref{eq:Hlike}) is maximized to obtain the maximum h-likelihood estimators (MHLEs) of ${\theta} = ({\beta}, {\alpha})$ and ${v} = ({v}_{\beta}, {v}_{\alpha})$. The score functions are given by 
\begin{align*}
	\begin{split}
		&{\partial \textit{h}}/{\partial {\beta}} = {X}^T {U}_\beta,\\ 
		&{\partial \textit{h}}/{\partial {\alpha}} = {X}^T {U}_\alpha,\\ 
		&{\partial \textit{h}}/{\partial {v}_{\beta}} = {Z}^T {U}_{\beta} - {U}_{v_\beta} \ \ \text{and}\\ 
		&{\partial \textit{h}}/{\partial {v}_{\alpha}} = {Z}^T {U}_\alpha - {U}_{v_\alpha}.
	\end{split}
\end{align*}
${X}$ is an $n\times p$ matrix whose $ij$th row is ${x}_{ij}$, ${Z}$ is an $n \times q$ matrix whose $ij$th row is ${z}_{ij}$, a vector indicating the cluster effect; ${U}_{\beta}$ and ${U}_{\alpha}$ are vectors of length $n$ such that 
\begin{align*}
	\begin{split}
		&U_{\beta ij} = \delta_{ij} - \tau_{ij} \Lambda_0(t_{ij}^{\gamma_{ij}})\ \ \text{and}\\ 
		&U_{\alpha ij} = \delta_{ij} + \bigg\{ \delta_{ij}\bigg(1 + \frac{t_{ij}^{\gamma_{ij}} \lambda_0'(t_{ij}^{\gamma_{ij}})}{\lambda_0(t_{ij}^{\gamma_{ij}})}\bigg) - \tau_{ij} t_{ij}^{\gamma_{ij}} \lambda_0(t_{ij}^{\gamma_{ij}})\bigg\}\gamma_{ij}\log{t_{ij}};\\
	\end{split}
\end{align*}
and ${U}_{v_\beta}$ and ${U}_{v_\alpha}$ are vectors of length $q$, such that 
\begin{align*}
	\begin{split}
		&U_{v_{\beta i}} = \frac{\partial \ell_2}{\partial v_{\beta i}} = \frac{1}{(1-\rho^2)}\bigg[\frac{v_{\beta i}}{\sigma_{\beta}^2} - \frac{\rho v_{\alpha i}}{\sigma_{\beta} \sigma_{\alpha}}\bigg]\ \ \text{and}\\
		&U_{v_{\alpha i}} = \frac{\partial \ell_2}{\partial v_{\alpha i}} = \frac{1}{(1-\rho^2)}\bigg[\frac{v_{\alpha i}}{\sigma_{\alpha}^2} - \frac{\rho v_{\beta i}}{\sigma_{\beta} \sigma_{\alpha}}\bigg].\\ 
	\end{split}
\end{align*}
The corresponding observed information matrix can be written explicitly as 
{\footnotesize
	\begin{align} \label{eq:H}
		{H}
		= -\frac{\partial^2 h}{\partial({\theta}, {v})^2}
		=   \begin{pmatrix}
			{X}^{T} {W}_{\beta} {X} & {X}^{T} {W}_{\beta\alpha} {X}& {X}^{T} {W}_{\beta} {Z} & {X}^{T} {W}_{\beta\alpha} {Z}\\
			{X}^{T} {W}_{\beta\alpha} {X} & {X}^{T} {W}_{\alpha} {X} & {X}^{T} {W}_{\beta\alpha} {Z} & {X}^{T} {W}_{\alpha} {Z} \\
			{Z}^T {W}_{\beta} {X} & {Z}^T {W}_{\beta\alpha} {X} & {Z}^T {W}_{\beta} {Z} + {Q}_{\beta}& {Z}^T {W}_{\beta\alpha} {Z} + {Q}_{{\beta}{\alpha}}\\
			{Z}^T {W}_{\beta\alpha} {X} & {Z}^T {W}_{\alpha} {X} & {Z}^T {W}_{\beta\alpha} {Z} + {Q}_{\beta\alpha} & {Z}^T {W}_{\alpha} {Z} + {Q}_{\alpha}
		\end{pmatrix},
\end{align}}
where ${X}$ and ${Z}$ are $n \times p$ and $n \times q$ model matrices for ${\beta}$, ${\alpha}$ and ${v}$ whose $ij$th rows are $x_{ij}$ and $z_{ij}$, respectively. ${W}_\beta$, ${W}_\alpha$ and ${W}_{\beta\alpha}$ are $n\times n$ diagonal matrices whose $ij$th diagonal elements are given by 
\begin{align*}
	\begin{split}
		w_{\beta ij} =\ & \tau_{ij} \Lambda_0(t_{ij}^{\gamma_{ij}}),\\ 
		w_{\alpha ij} =\ & -\bigg\{\delta_{ij}(1 + t_{ij}^{\gamma_{ij}}a_{ij}) + 
		\gamma_{ij} t_{ij}^{\gamma_{ij}} \log{t_{ij}} \bigg({\delta_{ij}}\bigg(\frac{t_{ij}^{\gamma_{ij}} \lambda_0''(t_{ij}^{\gamma_{ij}})}{\lambda_0(t_{ij}^{\gamma_{ij}})} + a_{ij} - 
		t_{ij}^{\gamma_{ij}}(a_{ij})^2 \bigg) - \\
		&\tau_{ij}( \lambda_0(t_{ij}^{\gamma_{ij}}) + t_{ij}^{\gamma_{ij}} \lambda_0'(t_{ij}^{\gamma_{ij}})) \bigg) - \tau_{ij}t_{ij}^{\gamma_{ij}}\lambda_0(t_{ij}^{\gamma_{ij}}) \bigg\} \gamma_{ij}\log{t_{ij}},\\ 
		w_{\beta\alpha ij} =\ & \tau_{ij} \gamma_{ij} t_{ij}^{\gamma_{ij}} \log{(t_{ij})} \lambda_0(t_{ij}^{\gamma_{ij}}),
	\end{split}
\end{align*}
where $a_{ij} = \frac{\lambda_0'(t_{ij}^{\gamma_{ij}})}{\lambda_0(t_{ij}^{\gamma_{ij}})}$. ${Q}_{\beta},\ {Q}_{\alpha}$, and ${Q}_{\beta\alpha}$ are $q \times q$ matrices whose $ij$th elements arise from $-{\partial^2 \ell_{2}}/{\partial {v}\partial {v}^T}$ such that 
\begin{align*}
	\begin{split}
		&{Q}_{\beta} = -\frac{\partial^2 \ell_{2}}{\partial {v}_\beta \partial {v}_{\beta}^T} = {{I}_q} \times \frac{1}{(1-\rho^2)\sigma_{\beta}^2},\\ 
		&{Q}_{\alpha} = -\frac{\partial^2 \ell_{2}}{\partial {v}_\alpha \partial {v}_{\alpha}^T} = {{I}_q}\times\frac{1}{(1-\rho^2)\sigma_{\alpha}^2},\\
		&{Q}_{\beta\alpha} = -\frac{\partial^2 \ell_{2}}{\partial {v}_\beta \partial {v}_{\alpha}^T} =-\frac{\partial^2 \ell_{2}}{\partial {v}_\alpha \partial {v}_{\beta}^T} = {I}_q \times -\frac{\rho}{(1-\rho^2)\sigma_{\beta}\sigma_{\alpha}},
	\end{split}
\end{align*}
where ${I}_q$ is a $q \times q$ identity matrix. Given ${\Sigma}$, the following system of Newton-Raphson equations can be solved iteratively for the MHLEs of ${\theta}^{(m + 1)} = ({\beta}^{(m + 1)^T}, {\alpha}^{(m + 1)^T})^T$ and ${v}^{(m+1)} = ({v}_{\alpha}^{(m + 1)^T}, {v}_{\beta}^{(m + 1)^T})^T$
{\footnotesize
	\begin{align} \label{eq:NR-BVN}
		\begin{pmatrix}
			\hat{{\theta}}^{(m+1)}\\
			\hat{{v}}^{(m+1)}
		\end{pmatrix} 
		=
		\begin{pmatrix}
			\hat{{\theta}}^{(m)}\\
			\hat{{v}}^{(m)}
		\end{pmatrix}
		+
		H^{-1}
		\begin{pmatrix}
			{\partial \textit{h}}/{\partial {\theta}}\\
			{\partial \textit{h}}/{\partial {v}} 
		\end{pmatrix} \Bigg|_{({\theta}, {v}) = (\hat{{\theta}}^{(m)}, \hat{{v}}^{(m)})},
\end{align}}
where $	{\partial \textit{h}}/{\partial {\theta}} = ({\partial \textit{h}}/{\partial{\beta}},\ 	{\partial \textit{h}}/{\partial{\alpha}})^T$ and ${\partial \textit{h}}/{\partial {v}} = (	{\partial \textit{h}}/{\partial {v}_\beta},\ 	{\partial \textit{h}}/{\partial {v}_\alpha})^T$.
For the estimation of the dispersion parameters, ${\Sigma} = (\sigma_\beta,\ \sigma_\alpha,\ \rho)^T$, the adjusted profile likelihood, $\textit{p}_{{\theta},{v}}(\textit{h})$, which eliminates $({\theta},{v})$ is used. Given $\hat{{\theta}} = (\hat{{\beta}}({\Sigma}), \hat{{\alpha}}({\Sigma}))$, $\hat{{v}} = (\hat{{v}}_{\beta}({\Sigma}), \hat{{v}}_{\alpha}({\Sigma}))$, $h$ given in (\ref{eq:Hlike}) and $ H$ defined in (\ref{eq:H}), $\textit{p}_{{\theta}, {v}}(\textit{h})$ is defined as follows:
\begin{equation}\label{eq:profileBVN}
	p_{{\theta},{v}}(\textit{h}) = \bigg[h - \frac{1}{2} \log \text{det}\{{H}/(2\pi)\} \bigg]\bigg|_{({\theta},{v}) = (\hat{{\theta}}, \hat{{v}})}.
\end{equation}

Solving the equations $\partial \textit{p}_{{\theta},{v}}(\textit{h})/\partial {\Sigma} = 0$ yields the restricted maximum likelihood estimators (REMLEs) of $\Sigma$. We opt for a non-linear optimizer implemented in \texttt{R} using the function \texttt{nlm}. The procedure iterates between $(\hat{{\theta}}, \hat{{v}})$ and $\hat{{\Sigma}}$ until all the estimates converge. The standard errors for $(\hat{{\theta}}, \hat{{v}})$ and $\hat{{\Sigma}}$ can be estimated directly from the inverse of the observed information matrices, ${H}$ and $-{\partial^2 p_{{\theta},{v}}(\textit{h})}/{\partial({\Sigma})^2}$ respectively \citep{ha2016interval,do2017statistical}.
\subsection{Fitting Algorithm}\label{sec:algorithm}
The model estimation algorithm described above can be summarized as follows:
\begin{description}
	\item  \textit{Initialisation:} 
	Using 0.01 as the initial values for the scale and shape coefficients, we fit a fixed effects Weibull MPR model. Estimates from this model are used as the initial values for the fixed parameters, ${\theta}$, in the mixed effects Weibull MPR model. For the initial values of the random effects, $v$, we use 0.01 and (0.1, 0.1, 0.1) is used for the dispersion parameters, ${\Sigma} = (\sigma_\beta, \sigma_\alpha, \rho)^T$.
	\item  \textit{Parameter Estimation:}
	\begin{description}
		\item \textbf{Step 1} Keeping the frailty variance parameters $\hat{{\Sigma}}$ fixed, maximize $h$ by iteratively re-solving the system of equations given in (\ref{eq:NR-BVN}) to obtain the new estimates $(\hat{{\theta}}, \hat{{v}})$.
		\item \textbf{Step 2} Given the estimates $(\hat{{\theta}}, \hat{{v}})$ from Step 1, a new estimate $\hat{{\Sigma}}$ is obtained by maximising (\ref{eq:profileBVN}) using \texttt{nlm}.
		\item Iterate between Step 1 and Step 2 until  the convergence criterion is met, that is until the maximum absolute difference between the previous and current estimates for $({\theta},{v})$ and $({\Sigma})$ is less than $10^{-6}$. After convergence is reached, the standard errors are estimated.
	\end{description}
	From the simulations we have carried out, we have found this algorithm to be time efficient and insensitive to starting values.
\end{description}
\section{Simulation Studies} \label{sec:sim}
The performance of the proposed methods is evaluated through simulation studies. The Weibull distribution is one of the most commonly used distributions in survival analysis, hence, we chose to generate the survival times from a Weibull MPR frailty model with the following regression parameters 
\begin{align*}
	\begin{split}
		{\beta} =& (\beta_0, \beta_1, \beta_2)^T = (1, -0.5, 0.5)^{T}\ \ \text{and}\\
		{\alpha} =& (\alpha_0, \alpha_1, \alpha_2)^T = (0.5, 0.5, -0.5)^{T}.
	\end{split}
\end{align*}
The corresponding covariates, ${x} = (1,x_{1},x_{2})^T$, were generated from an AR(1) process with a correlation coefficient of $0.5$. Each variable is marginally standard normal. The corresponding censored times were generated from a uniform distribution with a censoring rate of approximately $25\%$ or $50\%$ respectively. Following the two real data structures in Section \ref{sec:Data}, three different cluster sizes, $n_i \in (5, 20, 50)$, and two different cluster numbers, $q_i \in(20, 100)$, are considered (note that in contrast to this scheme, the real data has varying cluster sizes and this is something we have investigated, see Appendix). The frailty terms are generated from a bivariate normal distribution with the combination of various dispersion parameter values $\sigma_\beta \in (0.5, 1, 2)$, $\sigma_\alpha \in (0.25, 0.5, 1)$, $\rho \in (-0.5, 0.5)$. Each simulation scenario was replicated 500 times.

To summarize the simulation results, we compute the mean, the standard deviation (SE) and the average standard errors (SEE) for all the parameters we estimated for each scenario. The results for a censoring rate of $25\%$,  $\sigma_\beta = 1$, $\sigma_\alpha = 0.5$ and $\rho = -0.5$ are presented in Table \ref{tab:SimResults2}. Similar tables of the results from other simulation set-ups can be found in the Appendix. Overall, the h-likelihood estimates of both the fixed parameters and the frailty dispersion parameters perform quite well. The bias in the estimates is reduced as we increase both the cluster size and the number of clusters, this is observed in all the combinations of dispersion parameters and for both censoring percentages considered. The standard errors appear to be underestimated in the smaller sample sizes. This is especially true for the frailty variance parameters, and to a lesser extent, for the fixed effects. As we increase the cluster size and the number of clusters, however, we see that the standard errors shrink for all the parameters, and the SEE converges towards the SE; in any case, the use of the standard errors in hypothesis testing for the frailty variance parameters is recommended against \citep{maller2003testing}. Similar but larger bias is observed in the results of scenarios with a $50\%$ censoring (see Appendix).
\begin{table}[ht]
	\caption{Averaged coefficient estimates, standard deviations (SE) and the average standard errors (SEE) for the simulation scenario with dispersion parameters $\sigma_\beta = 1$, $\sigma_\alpha = 0.5$ and $\rho = -0.5$ and a censoring rate of $25\%$.} \label{tab:SimResults2}
	\footnotesize
	\begin{center}
		\begin{tabular}{||lccccccccc||} 
			\hline
			& $\hat{\beta}_0$ & $\hat{\beta}_1$ & $\hat{\beta}_2$ & $\hat{\alpha}_0$ & $\hat{\alpha}_1$ & $\hat{\alpha}_2$ & $\hat{\sigma_\beta}$ & $\hat{\sigma_\alpha}$ & $\hat{\rho}$\\	
			& Mean  & Mean  & Mean & Mean & Mean & Mean  & Mean  & Mean & Mean \\
			&  (SE) & (SE) & (SE) & (SE) & (SE) & (SE) & (SE) & (SE) & (SE) \\
			($q_i, n_i$)&  (SEE) & (SEE) & (SEE) & (SEE) & (SEE) & (SEE) & (SEE) & (SEE) & (SEE) \\
			True & 1 & -0.5 & 0.5 & 0.5 & 0.5 & -0.5 & 1 & 0.5 & -0.5 \\ 
			\hline
			\hline
			(20, 5) & 1.30 & -0.58 & 0.59 & 0.64 & 0.50 & -0.49 & 1.18 & 0.51 & -0.41 \\ 
			& (0.35) & (0.27) & (0.25)  & (0.18)  & (0.11)  & (0.11)  & (0.33)  & (0.15)  & (0.39)  \\ 
			& (0.32) &  (0.21) & (0.20) & (0.15) & (0.09) & (0.09) & (0.20) & (0.09) & (0.19) \\
			
			(20, 20) & 1.05 & -0.52 & 0.52 & 0.54 & 0.50 & -0.50 & 1.00 & 0.50 & -0.50 \\ 
			& (0.24) & (0.08) & (0.08) & (0.12) & (0.04)  & (0.04) & (0.20)  & (0.09) & (0.23)\\ 
			& (0.24) & (0.08) & (0.08) & (0.12) & (0.04) & (0.04) & (0.16) & (0.08) & (0.17)\\
			
			(20, 50) & 1.03 & -0.51 & 0.50 & 0.52 & 0.50 & -0.50 & 1.01 & 0.50 & -0.49 \\ 
			& (0.23) & (0.05) & (0.05) & (0.12) & (0.02) & (0.02) & (0.17) & (0.08) & (0.19) \\ 
			& (0.23) & (0.05) & (0.05) & (0.12) & (0.02) & (0.02) & (0.16) & (0.08) & (0.17) \\
			
			(100, 5) & 1.21 & -0.56 & 0.57 & 0.59 & 0.49 & -0.49 & 1.07 & 0.52 & -0.46 \\ 
			& (0.15) & (0.10) & (0.10) & (0.08) & (0.04) & (0.04) & (0.13) & (0.06) & (0.16) \\ 
			& (0.13) &  (0.08) & (0.08) & (0.07) & (0.04) & (0.04)  & (0.08) &  (0.04) & (0.09)  \\
			
			(100, 20) & 1.05 & -0.51 & 0.51 & 0.53 & 0.50 & -0.50 & 1.01 & 0.50 & -0.50 \\ 
			& (0.11) & (0.04) & (0.04) & (0.05) & (0.02) & (0.02) & (0.09) & (0.04) & (0.10)  \\ 
			& (0.11) & (0.04) & (0.04) & (0.05) & (0.02) & (0.02) & (0.07) & (0.04) & (0.08)\\
			
			(100, 50) & 1.02 & -0.50 & 0.50 & 0.51 & 0.50 & -0.50 & 1.00 & 0.50 & -0.50 \\ 
			& (0.11) & (0.02) & (0.02) & (0.05) & (0.01) & (0.01) & (0.08) & (0.04) & (0.08) \\ 
			& (0.10) & (0.02) & (0.02) & (0.05) & (0.01) & (0.01) & (0.07) & (0.04) & (0.08) \\
			\hline
		\end{tabular}
	\end{center}
\end{table}
\newpage
\section{Data Analysis} \label{sec:Data}
\subsection{Modelling Details} \label{sec:data_models}
We also illustrate the proposed models on two datasets. Though our main focus is the MPR model with BVN frailties, for the purpose of comparison and analsyis we also consider various simplifications of this model. More precisely, we fit Weibull MPR models with the following frailty structures to each of the datasets we consider: 
\begin{itemize}
	\itemsep0em 
	\item BVNF: BVN frailty ($-1 \le \rho \le 1$)
	\item IF: independent frailty ($\rho = 0$)
	\item CF: common frailty ($\rho = \pm1$)
	\item ScF: scale frailty only ($\sigma^2_\alpha = 0$), i.e., the classical multiplicative frailty model
	\item ShF: shape frailty only ($\sigma^2_\beta = 0$)
\end{itemize}

These models are fitted following the procedures described in Sections \ref{sec:estimation} and \ref{sec:algorithm}, and the standard errors for the estimated parameters are computed as described in Section \ref{sec:estimation}. For the selection of the fraily structure that is best supported by the data, we use the Akaike information criterion (AIC). Various extended definitions of the AIC in random-effect models can be formulated based on different likelihood functions \citep{vaida2005conditional, xu2009using, ha2007model, do2017statistical}. More specifically, we make use of the restricted AIC (rAIC) \citep{ha2007model} and the conditional AIC (cAIC) \citep{vaida2005conditional, do2017statistical}. The rAIC is based on the restricted likelihood approximation $p_{{\theta},{v}}(h)$, which eliminates $({\theta},{v})$ from $h$ and thus is a function of the frailty parameters only. While the cAIC is based on the conditional joint density function for $t_{ij}$ and $\delta_{ij}$ given the random effects, $\ell_{1ij}$. Definitions of the AICs are as follows:
$$ rAIC = -2 p_{{\theta},{v}}(h) + 2df_r,$$ 
$$ cAIC = -2 \sum_{ij}\ell_{1ij} + 2df_c,$$ 
where $df_r$ is the number of dispersion parameters governing the frailty distribution, $df_c = \text{trace}({H}^{-1}{H}^{*})$, the effective degrees of freedom adjusted for the fixed and random effect estimates, and ${H}^{*} = -\partial^2 \sum\ell_{1ij}/ \partial ({\theta},{v})^2$ \citep{do2017statistical, lee2017generalized}. The computation of $df_c$ involves the fixed effects, the random effects and the frailty distribution parameters, but in a model with no frailty, $df_c$ is just the number of fixed effects in the model, and hence the cAIC becomes the classical AIC in this case \citep{do2017statistical}. After fitting each of the aforementioned models as well as a model with no frailty (NF), we obtain the corresponding rAIC and cAIC values for the purpose of model comparison. 

In the Weibull MPR models (or any other parametric MPR model with a scale and shape parameter), the scale coefficients describe the overall scale of the hazard and the shape coefficients describe its evolution over time. A positive scale coefficient indicates an increase in the hazard relative to some reference category and similarly, a positive shape coefficient indicates an increased hazard over time relative to some reference category. While an examination of the ${\beta}$ and ${\alpha}$ coefficients separately provides some initial understanding of the effect of a variable; it is important to look at the combined information from both coefficients when determining its overall effect, and hence, it is important to look at the hazard ratios. For a binary covariate, $x_k$, \citet{burke2017multi} show the hazard ratio under the Weibull MPR frailty model is given by $$\text{HR}_{k}(t) = \frac{\lambda(t|x_k = 1)}{\lambda(t|x_k = 0)} = \exp(\beta_k + \alpha_k) t^{\exp({x}_{(-k)}^{T} {\alpha})\{\exp(\alpha_k) - 1\}},$$ where $\beta_k$ and $\alpha_k$ are, respectively, the scale and shape coefficients of $x_k$, and ${x}_{(-k)} = (1, x_1, \dots, x_{k-1}, 0, x_{k + 1}, \dots, x_p)$, the covariate vector with $x_k$ set to 0. (Note here that we have dropped the subscripts $ij$ for notational convenience.) Because $\text{HR}_{k}$ depends on the values of the other covariates in the model via the vector ${x}_{(-k)}$, we set them to their empirical modal values. In line with this, we also set the random effect $v_\alpha$ to its modal value of zero. 
\subsection{Extensive-Stage Small-Cell Lung Cancer}
This dataset was collected as part of a randomized, multi-centre study conducted by the Eastern Cooperative Oncology Group (ECOG). The main purpose of the trial was to determine if cyclic alternating combination chemotherapy was superior to cyclic standard chemotherapy in patients with extensive-stage small-cell lung cancer. Patients were randomly assigned to one of two treatment arms, standard chemotherapy (CAV; reference category) or an alternating regimen (CAV-HEM). The dataset includes 579 patients from 31 different institutions, with the number of patients per institution ranging from 1 to 56 and a median of 17 patients per institution. The outcome variable was time (in years) from randomisation until death. The median survival time and maximum follow-up time were 0.86 years and 8.45 years, respectively, and of the 579 study participants only 10 were censored yielding a censoring rate of approximately 1.7\%. Besides the survival time, censoring status, institution code and treatment, four other dichotomous variables were included in this dataset, namely (reference category listed first): the presence of bone metastases (no, yes), the presence of liver metastases (no, yes), patient status on entry (confined to bed or chair, ambulatory), and whether there was a weight loss prior to entry (no, yes). More details on the trial and it's clinical results can be found in \citet{ettinger1990randomized}. This dataset was also previously analysed in \citet{gray1994bayesian} using a fully Bayesian approach, in \citet{vaida2000proportional} using a marginal likelihood approach, and in \citet{ha2016interval} using a correlated frailty model fitted using a h-likelihood approach.

We fit the models listed in Section \ref{sec:data_models} to this lung cancer dataset and the results are presented in Table \ref{tab:LC_ECOG1}. To explore the degree of dependence between the two random components, ${v}_\beta$ and ${v}_\alpha$, we first fit the BVNF model, (i.e., the model which assumes a bivariate normal distribution). The estimate of $\rho$ ($\hat{\rho} = 0.995$) indicates a very strong positive correlation between the predicted random components $\hat{{v}}_\beta$ and $\hat{{v}}_\alpha$. This perhaps suggests that the model could be simplified and only one random component is needed. Note that the large $\hat{\phi}$ value is due to the very small $\hat{{v}}_\beta$ values, and this may be pointing towards a shape frailty only model.

\begin{table}[!htbp]
	\caption{The coefficient estimates, frailty dispersion parameter estimates and estimated standard errors (in brackets) from each model we fit to the lung cancer dataset.}
	\label{tab:LC_ECOG1}
	\footnotesize
	\begin{center}
		\begin{threeparttable}
		\begin{tabular}{||p{1.7cm}lcccccc||}
			\hline 
			& & NF & BVNF & IF & CF & ScF & ShF\\
			\hline 
			\hline
			\multirow{12}{*}{Scale}
			& Intercept & 0.11 & 0.12  &  0.11 & 0.11 & 0.12 & 0.11\\
			&  & (0.13) & (0.13) & (0.13) & (0.13) & (0.13) & (0.13)\\
			& Treatment  & -0.24 & -0.23 & -0.24 & -0.24 & -0.24 & -0.24\\ 
			& \hspace{2.5pt} CAV-HEM & (0.09) & (0.09) & (0.09) & (0.09) & (0.09) &  (0.09)\\
			& Bone Metastases  &  0.23 & 0.24 & 0.22 & 0.22 & 0.25 &  0.22 \\
			& \hspace{2.5pt} Yes & (0.10) & (0.10) & (0.10) & (0.10) & (0.10) & (0.10) \\
			& Liver Metastases &  0.33 & 0.38 & 0.39 & 0.39 & 0.32 & 0.39 \\
			& \hspace{2.5pt} Yes & (0.10) & (0.10) &  (0.10) & (0.10) & (0.10) & (0.10)\\
			& Patient Status & -0.58 & -0.63 & -0.62 & -0.62 & -0.60 & -0.62\\
			& \hspace{2.5pt} Ambulatory & (0.11)& (0.11) & (0.11) & (0.11) & (0.11) & (0.11)\\
			& Weight Loss &  0.19 &0.21  & 0.21 & 0.21 &  0.19 & 0.21 \\
			& \hspace{2.5pt} Yes & (0.10) & (0.10)& (0.10) & (0.10) & ( 0.10) & (0.10)\\
			\hline 
			\multirow{12}{*}{Shape} 
			& Intercept  & 0.14 & 0.23 &  0.22 &  0.22  &  0.16 & 0.22\\
			& & (0.08) & (0.12) & (0.12) & (0.12) & (0.08) & (0.12)\\
			& Treatment & -0.28 & -0.23 & -0.24 & -0.24  & -0.27 & -0.24 \\
			& \hspace{2.5pt} CAV-HEM & (0.06) & (0.07) & (0.07) &  (0.07) & (0.06) & (0.07)\\
			& Bone Metastases & 0.04 & 0.02 & 0.03 &  0.03 & 0.03 & 0.03\\
			& \hspace{2.5pt} Yes & (0.07)& (0.08) & (0.08) & (0.08) & (0.07)  & (0.08) \\
			& Liver Metastases & -0.14 & -0.11 & -0.11 & -0.11 & -0.13 & -0.11\\
			& \hspace{2.5pt} Yes & (0.07) & (0.07) &  (0.07) & (0.07) & (0.07) & (0.07)\\
			& Patient Status &  0.33 & 0.38 & 0.38 &  0.38 & 0.33 & 0.38\\
			& \hspace{2.5pt} Ambulatory & (0.07) & (0.09) & (0.09)& (0.09) & (0.08) & (0.09)\\
			& Weight Loss & 0.06 & 0.02 & 0.03 & 0.03 & 0.05 & 0.03\\ 
			& \hspace{2.5pt} Yes & (0.07)& (0.08) & (0.08) & (0.08) & (0.07) & (0.08)
			\\
			\hline 
			\multirow{8}{*}{\parbox{2cm}{Frailty\ \ \ \ \ \ \ parameters}}
			& $\hat{\sigma}_{\beta}$ & & 0.08 & 0.00 & 0.00  & 0.16 & \\
			& & & (0.01) &  (0.02)&  (0.00)  & (0.03)& \\
			& $\hat{\sigma}_{\alpha}$ & & 0.29 & 0.31 & & & 0.31 \\
			&  & & (0.04) &  (0.05) & &  & (0.05)\\
			& $\hat{\rho}$ & & 1.00 & & &  & \\
			&  & & (0.04) &  & & & \\
			& $\hat{\phi}$ & & & & 3666.42 & & \\
			& & & & & (439.09) & & \\
			\hline 
			$-2 p_{{\theta},{v}}(h)$ & & 1123.40 & 1077.74 & 1079.52 & 1079.52 & 1121.35 & 1079.52 \\ 
			$df_r$ & & 0 & 3 & 2 & 2 & 1 & 1 \\
			\multicolumn{2}{||l}{$rAIC - rAIC_{min}$} &  41.88 & 2.22 & 2.00 & 2.00 & 41.83 &  0 \\
			$df_c$ & & 12.00 & 30.96 & 30.89 & 30.89 & 19.89 & 30.89\\
			\multicolumn{2}{||l}{$cAIC - cAIC_{min}$} & 65.19 & 0 & 2.08 & 2.06 & 61.09 & 2.08 \\
			\hline 
		\end{tabular}

	{\footnotesize
		\begin{tablenotes}
			\item[]{Note that for the CF model, an estimate for $\sigma_\alpha$ can be found by evaluating $\hat{\phi}\hat{\sigma}_\beta$. Given the above values $\hat{\sigma}_\alpha = 0.31$.}
	\end{tablenotes}}
\end{threeparttable}
	\end{center}
\end{table}

With the exception of the ScF model, all the models containing at least one frailty component have quite similar rAIC values, all much smaller than the rAIC value from a no frailty model, with the model with a frailty term in the shape only, the ShF model, having the lowest value. The ScF model, the more standard multiplicative frailty model, has a similar rAIC value to the model with no frailty components, suggesting that the baseline risk is homogenous across the centres and a frailty term in the scale is not needed. We observe a similar pattern in the cAIC values, in that all the models containing at least one frailty component (with the exception of the ScF model) have similar cAIC values and all much smaller compared to the NF model. Although the BVNF model has the lowest cAIC value, it is still close to those of the IF, CF, and ShF models. Given this, and also the rAIC values, we consider the ShF as the ``best'' model.

Although we report the standard errors of the frailty variance parameters, one should not use them for testing the hypothesis $H_0: \sigma =  0$ \citep{vaida2000proportional}. A likelihood ratio test can be carried out instead. Since the value of $\sigma$ in the null hypothesis is at the boundary of the parameter space, the standard approximation of the loglikelihood ratio statistic by a $\chi_{1}^{2}$ distribution often leads to over conservative test results \citep{self1987asymptotic, stram1994variance, duchateau2007frailty}. To correct this bias, a mixture of a chi-square distribution with one and zero degrees of freedom, $(\chi_{0}^{2} + \chi_{1}^{2})/2$, should be used as the approximation of the log-likelihood ratio statistic \citep{maller2003testing}. The test statistic  at the 5\% significance level is thus 2.71. The difference in deviance, $-2 p_{{\theta},{v}}(h)$ between the NF model and the ShF model is 43.878, and hence the scale frailty is significant, i.e., $\sigma_{\alpha} > 0$. 

Focusing on the results from the model selected by the rAIC, the ShF model; considering the scale parameter results first, all the variables included in the model have significant scale coefficients. The CAV-HEM treatment and the subject being ambulatory on entry have negative scale coefficients and so reduce the hazard of death relative to their respective reference categories, the CAV treatment and subject being confined to bed or chair on entry respectively. The presence of bone metastases, the presence of liver metastases and weight loss prior to study entry are all found to increase the hazard of death relative to their reference categories. Now, considering the shape parameter, only two variables, treatment and patient status on entry have significant coefficients. The CAV-HEM treatment has a negative shape coefficient suggesting that the hazard further decreases over time, relative to the reference category. The positive shape coefficient for the subject being ambulatory on entry suggests that the hazard increases over time, although this variable has an initial effect of reducing the hazard of death, this effect wears off over time, relative to its reference category. 

\begin{figure}[!htbp]
	\begin{center}
		\includegraphics[scale= 0.85]{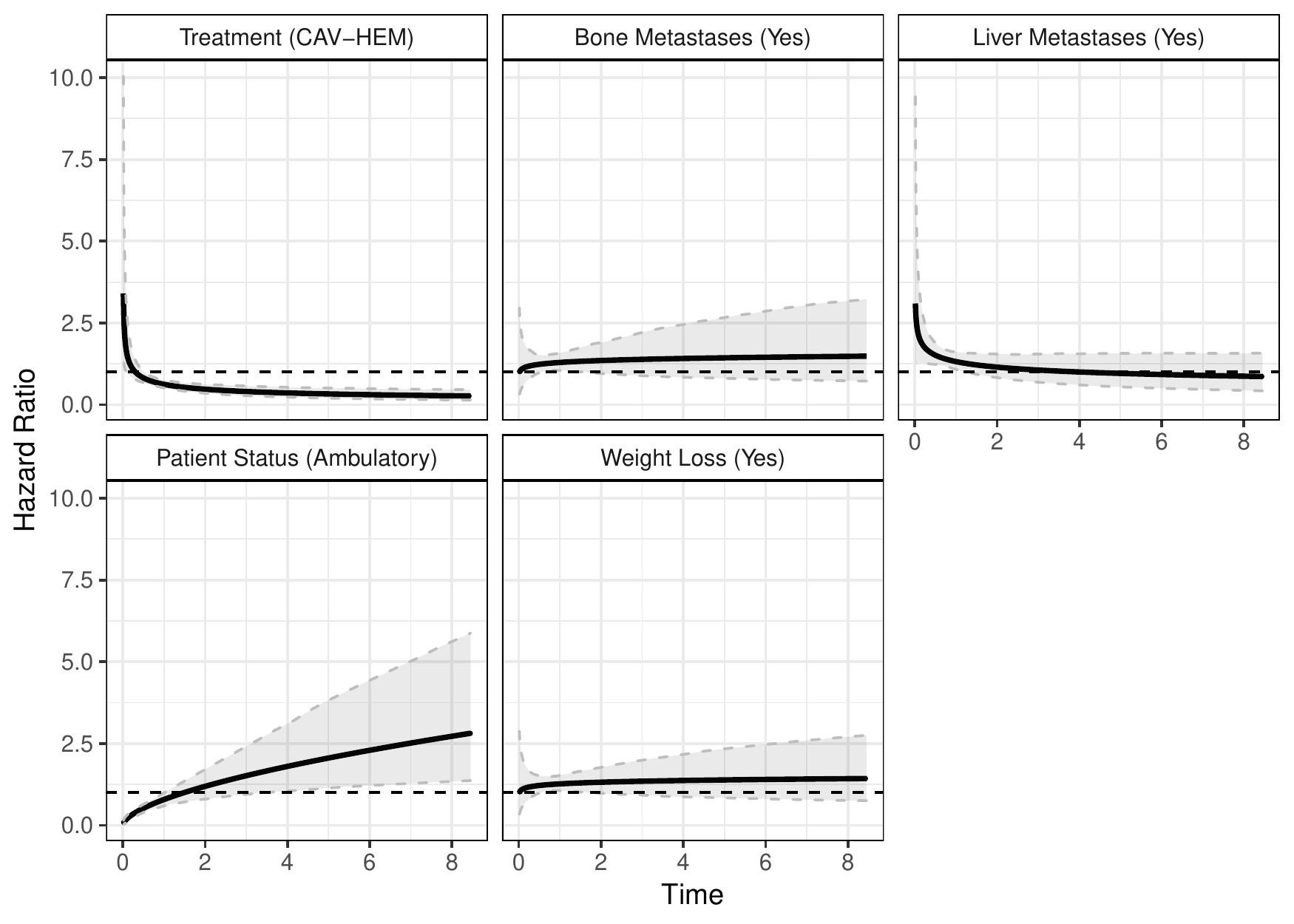}
	\end{center}
	\caption{The hazard ratios with 95\% confidence intervals based on results from the ShF model. The modal values were used in the computation of these hazard ratios (Treatment = CAV, Bone Metastases = no, Liver Metastases = no, Patient Status =  ambulatory on entry, Weight Loss = yes). \label{fig:M5_HR}}
\end{figure}

Figure \ref{fig:M5_HR} shows the hazard ratio corresponding to each of the variables in our model along with 95\% confidence intervals estimated using a parametric bootstrap \citep{davison1997bootstrap}. The presence of bone metastases and weight loss prior to study entry appear to have a more or less constant hazard over time, this is expected since their corresponding shape effects are quite small and not significant. The presence of liver metastases has a negative effect on the hazard but this effect wears off within the first 2 to 3 years. The hazard ratio for the patient being ambulatory on entry appears to be increasing over time. The effectiveness of the CAV-HEM treatment is only observed after 2 months or so from the treatment start date and the hazard continues to decrease over time relative to the CAV treatment.

\begin{figure}[!htbp] 
	\begin{center}
		\includegraphics[scale= 0.75]{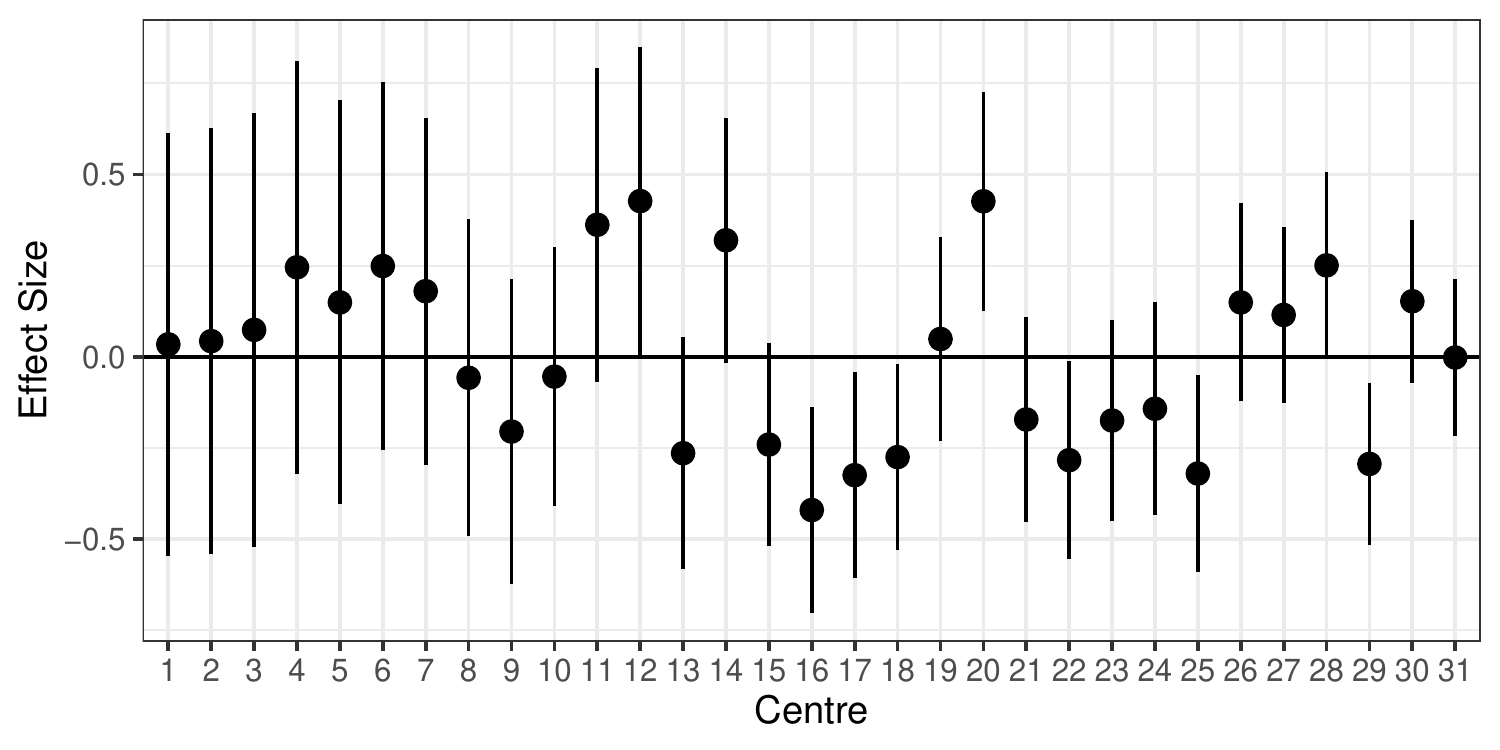}
	\end{center}
	\caption{The random effects of the 31 centres with 95\% confidence intervals under the ShF model. Centres are sorted in increasing order based on the number of patients. \label{fig:M5_valpha}}
\end{figure}

The random effects along with their 95\% confidence intervals under the ShF model are shown in Figure \ref{fig:M5_valpha}. Note that as the cluster size increases, the confidence bounds around the cluster effect shrink. The biggest changes in centre specific hazard over time can be seen in centres 12, 16 and 20. A positive frailty suggests the hazard is increasing over time relative to the baseline, while a negative one suggests it's decreasing relative to the baseline.

\newpage
\subsection{Bladder Cancer}
This multi-centre dataset was collected as part of the European Organisation for Research and Treatment of Cancer (EORTC) trial 30791 \citep{sylvester2006predicting}. A total of 410 patients with superficial bladder cancer were included in this dataset. The patients came from 21 different centres and the number of patients per centre varied between 3 and 78 patients with a median of 15 patients per centre. The outcome variable is relapse-free or disease-free interval after transurethral resection, i.e., time from randomisation until cancer relapse. Patients which did not experience a recurrence during the follow-up period were censored at their last date of follow-up. The maximum follow-up time was 10.16 years and of the 410 patients, 204 (approximately 50\% of the patients) were right-censored. The two covariates included in this dataset are (reference categories are listed first): a treatment indicator for chemotherapy (no, yes), and a variable representing the prior recurrence (no, yes). This dataset was also previously analysed in \citet{ha2011frailty} and can be found in the \texttt{R} package \texttt{frailtyHL} \citep{do2012frailtyhl, do2017statistical}. As in the previous example, we fit the models listed in Section \ref{sec:data_models} to this dataset and compare the fit using the rAIC and cAIC. The results are presented in Table \ref{tab:Blad1}.
\begin{table}[!htbp]
	\caption{The coefficient estimates, frailty dispersion parameter estimates and estimated standard errors (in brackets) from each model we fit to the bladder cancer dataset.}\label{tab:Blad1}
	\footnotesize
	\begin{center}
		\begin{threeparttable}
		\begin{tabular}{||p{1.7cm}lcccccc||}
			\hline 
			& & NF & BVNF & IF & CF & ScF & ShF \\
			\hline 
			\hline
			\multirow{6}{*}{Scale}
			& Intercept & -0.79 & -0.71 & -0.70 & -0.70 & -0.70 & -0.79\\
			& & (0.18)& (0.19) & (0.20) & (0.20) & (0.20) & (0.18)\\
			& Chemotherapy & -0.72 & -0.74 & -0.74 & -0.74 & -0.74 &  -0.72 \\ 
			& \hspace{2.5pt} Yes& (0.19) & (0.19) & (0.19) & (0.19) & (0.19) & (0.19)\\
			& Prior Recurrence & 0.55  & 0.57 & 0.57 & 0.57 & 0.57 &  0.55 \\
			& \hspace{2.5pt} Yes& (0.17) & (0.17) & (0.17) & (0.17) & (0.17) & (0.17)\\
			\hline 
			\multirow{6}{*}{Shape}
			& Intercept & -0.19 & -0.17 & -0.19 & -0.19 & -0.19 & -0.19 \\
			& & (0.13) & (0.13) & (0.13) & (0.13) & (0.13) & (0.13)\\
			& Chemotherapy & 0.03 & 0.02 & 0.03 & 0.03 & 0.03 &  0.03\\ 
			& \hspace{2.5pt} Yes& (0.13) & (0.13) & (0.13) & (0.13) & (0.13) & (0.13)\\
			& Prior Recurrence & -0.01 & 0.01 & 0.02 & 0.02 &  0.02 &  -0.01\\
			& \hspace{2.5pt} Yes& (0.12) & (0.12) & (0.12) & (0.12) & (0.12) & (0.12)\\
			\hline 
			\multirow{4}{*}{\parbox{2cm}{Frailty\ \ \ \ \ \ \ parameters}}
			& $\hat{\sigma}_{\beta}$ & & 0.22 & 0.28 & 0.27 & 0.28 & \\
			& &  & (0.06) & (0.06) & (0.06) & (0.06) & \\
			& $\hat{\sigma}_{\alpha}$ & & 0.06 &  0.00 & & & 0.03\\
			& & & (0.02) & (0.03) & & & (0.03)\\
			& $\hat{\rho}$ & & 1.00 & & & & \\
			& & & (0.07) & & & &\\
			& $\hat{\phi}$ & & & & 0.07 & & \\
			& & & & & (0.22) & & \\
			\hline 
			$-2 p_{{\theta},{v}}(h)$ & & 946.96 & 943.75 & 943.28 & 943.28 & 943.28 & 946.96\\
			$df_r$ & & 0 & 3 & 2 & 2 & 1 & 1 \\
			\multicolumn{2}{||l}{$rAIC - rAIC_{min}$} &  1.68 & 4.47 & 2.00 & 2.00 & 0 & 3.68 \\ 
			$df_c$ & & 6.00 & 12.76 & 13.09 & 13.11 & 13.09 & 6.35 \\
			\multicolumn{2}{||l}{$cAIC - cAIC_{min}$} & 6.46 & 0.70 & 0.05 & 0 & 0.05 & 6.45\\
			\hline 
		\end{tabular}	
	{\footnotesize\begin{tablenotes}
			\item[]{Note that for the CF model, an estimate for $\sigma_\alpha$ can be found by evaluating $\hat{\phi}\hat{\sigma}_\beta$. Given the above values $\hat{\sigma}_\alpha = 0.02$.}
	\end{tablenotes}}
\end{threeparttable}
	\end{center}
\end{table}

Similar to the previous example, the correlation coefficient between the two random effects, ${v}_\beta$ and ${v}_\alpha$ is $\approx 1$ suggesting that the model can be simplified. The model with the single frailty parameter in the scale has the lowest rAIC value and therefore is the preferred model, suggesting that there is substantial variation in the baseline risk across the centres and this variation is constant overtime. All the models containing a scale frailty term have similar cAIC values, which are significantly lower than the values for the NF model and the ShF model, suggesting the need for a scale frailty term. Since the difference in cAIC between the models BVNF, IF, CF and ScF is small, we focus on the simplest model in that set, the ScF model.

\begin{figure}[!htbp]
	\begin{center}
		\includegraphics[scale= 0.75]{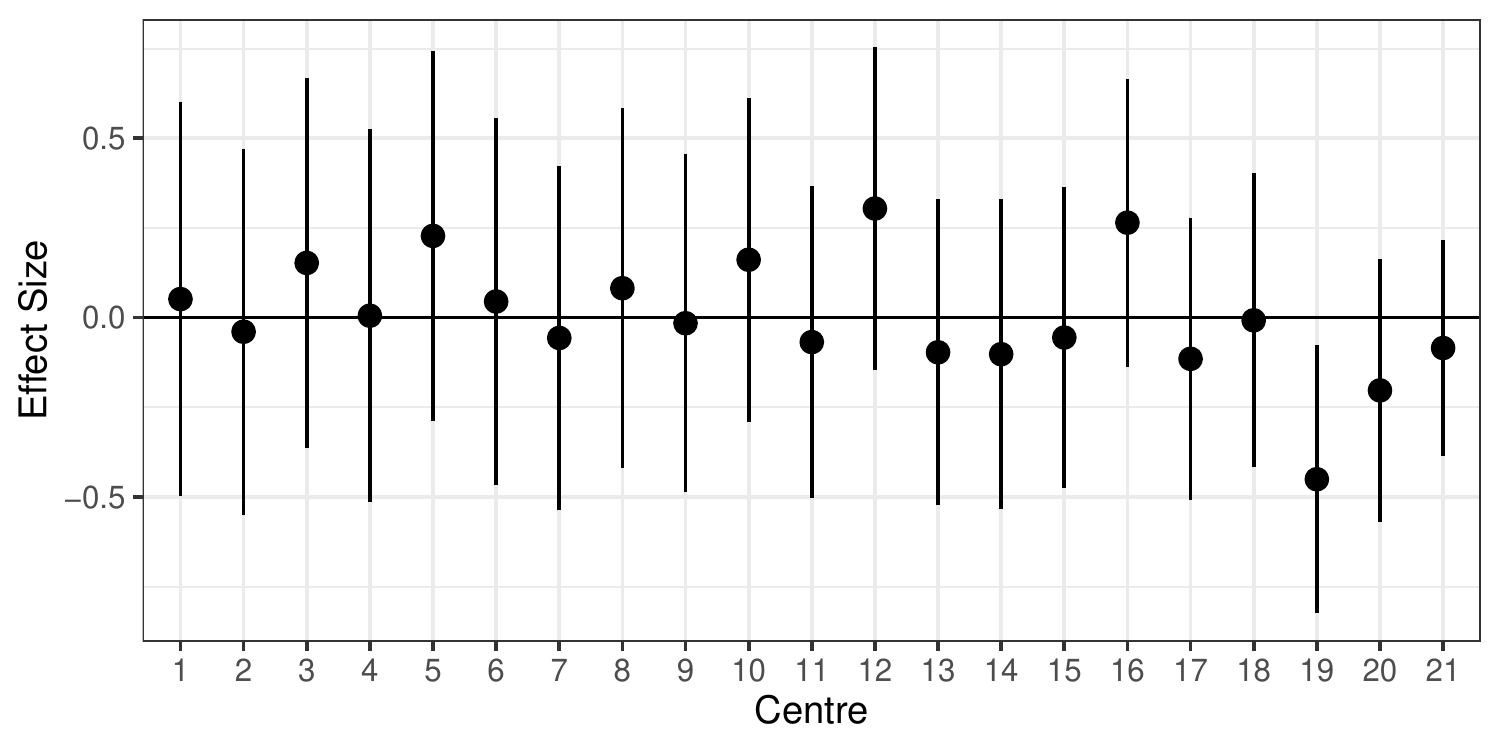}
	\end{center}
	\caption{The random effects of the 21 centres with 95\% confidence intervals under the ScF model. Centres are sorted in increasing order based on the number of patients. \label{fig:M4_valpha_bladder}} 
\end{figure}

The difference in deviance, $-2 p_{{\theta},{v}}(h)$, between the NF model and the ScF model is $3.682$ ($>2.71$), and hence the scale centre effect is significant at the $5\%$ significance level, i.e., $\sigma_{\beta} > 0$. A caterpillar plot of the random effects from the ScF model is presented in Figure \ref{fig:M4_valpha_bladder} and again the centres are sorted by the number of patients. Centre 19 appears to have the only significant effect; and hence, the observed frailty effect is solely due to this one centre having a lower hazard than the other centres included in the study. This, perhaps, explains why the rAIC is only slightly lower than the NF model. 

\begin{figure}[!htbp] 
	\begin{center}
		\includegraphics[scale= 0.75]{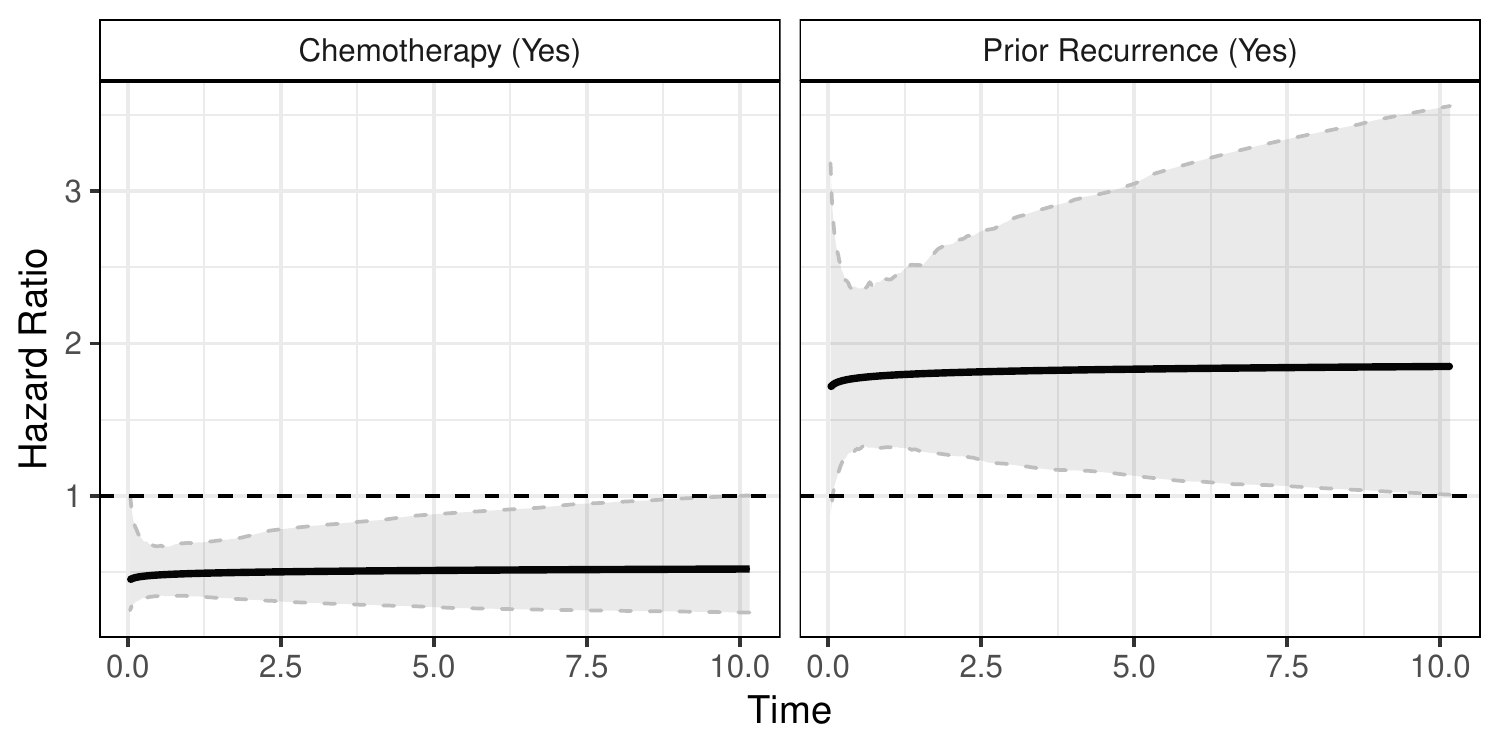}
	\end{center}
	\caption{The hazard ratios with 95\% confidence intervals based on the ScF model. The modal values were used in the computation of these hazard ratios (Chemotherapy = yes, Prior Recurrence = no). \label{fig:M4_HR}}
\end{figure}

Focusing on the results from the ScF model, both variables included in the model have significant scale coefficients but non-significant shape coefficients. The variable Chemotherapy has a negative coefficient and so the hazard is significantly reduced, time until recurrence is prolonged for patients that received chemotherapy relative to those who did not. Prior Recurrence has a positive coefficient and so having had a recurrence already significantly increases the risk of another recurrence, relative to it being a primary occurrence. The shape coefficients corresponding to the two variables, albeit non-significant, are both positive, suggesting that the effect of chemotherapy wears off over time, while the longer one survives after a recurrence the higher the risk of another recurrence. Plots of the hazard ratios along with their bootstrapped 95\% confidence intervals can be seen in Figure \ref{fig:M4_HR}. As expected, the two variables have approximately constant hazards over time and so perhaps the model can be reduced to a proportional hazards model.   

\newpage
\section{Discussion and Conclusions}	
The MPR frailty modelling framework we have proposed not only includes frailty structures not previously explored in the literature, but also generalizes a variety of existing sub-models. Existing literature on MPR frailty models has been limited to multiplicative frailty, and, to the best of our knowledge, a model with correlated frailty in each distributional parameter has not previously been considered. We believe that this is a natural structure in the context of MPR modelling, since it has been shown that estimates of the scale and shape can be quite correlated in practice \citep{burke2017multi}; hence, it is useful to allow for the possibility that correlation may propagate to the frailty terms.

Although the numerical studies were carried out on a Weibull MPR model, we have developed the model and the estimation procedure in a generic form; the underlying cumulative hazard function can be replaced with that of any other two parameter distribution. In principle, the methods can also be extended to models with more than two distributional parameters, e.g., the power generalized Weibull model of \citet{burke2020flexible}, using a higher order multivariate normal distribution but this is beyond the scope of this paper. The adopted h-likelihood framework provides a computationally inexpensive and straightforward two step procedure to fit our frailty models while avoiding the often intractable integration of the random effects over the frailty distribution. Moreover, the readily available estimates of the frailties allow for the survivor function for individuals with specific characteristics to be estimated, and this is useful in providing information about the merits of the different centres in terms of patient survival in multi-centre studies.

While the proposed MPR framework provides a very flexible approach to modelling correlated survival data at a minimal computational cost, there are various ways in which we can extend it to handle more complicated frailty structures. All of the models that we have considered have a constant frailty variance, perhaps a natural next step for us is to allow the frailty variance to depend on covariates in a similar fashion to that of \citet{peng2020multiparameter}. Another potential direction worth exploring would be multilevel or nested frailty structures, we can have data on patients, nested within centres or hospitals, with recurrent event times and hence we may need a frailty component for the patients and a frailty component for the centre or hospital. The procedures we present can be straightforwardly extended to include more than one random component for each distributional parameter.   

\newpage
\section*{Acknowledgements}

The first author is funded by the Irish Research Council. The second author is supported by the National
Research Foundation of Korea (NRF) grant funded by the Korea government (MSIT) (No. NRF-2020R1F1A1A01056987).

\bibliographystyle{apalike}
\bibliography{MPR_hlike_bibliography}

\newpage
\centerline{\sc{Appendix}}
\medskip

\appendix

\section{Other Simulation Studies}

This section displays analogous simulation results to those of Table \ref{tab:SimResults2} from the main paper
but for different frailty dispersion parameters and/or censoring rate.

\begin{table}[!ht]
	\caption{Averaged coefficient estimates, standard deviations (SE) and the average standard errors (SEE) for the simulation scenario with dispersion parameters $\sigma_\beta = 2$, $\sigma_\alpha = 1$ and $\rho = -0.5$ for $25\%$ censoring rate.}
	\footnotesize
	\begin{center}
		\begin{tabular}{||lccccccccc||} 
			\hline
			& $\hat{\beta}_0$ & $\hat{\beta}_1$ & $\hat{\beta}_2$ & $\hat{\alpha}_0$ & $\hat{\alpha}_1$ & $\hat{\alpha}_2$ & $\hat{\sigma_\beta}$ & $\hat{\sigma_\alpha}$ & $\hat{\rho}$\\	
			& Mean  & Mean  & Mean & Mean & Mean & Mean  & Mean  & Mean & Mean \\
			&  (SE) & (SE) & (SE) & (SE) & (SE) & (SE) & (SE) & (SE) & (SE) \\
			($q_i, n_i$)&  (SEE) & (SEE) & (SEE) & (SEE) & (SEE) & (SEE) & (SEE) & (SEE) & (SEE) \\
			True & 1 & -0.5 & 0.5 & 0.5 & 0.5 & -0.5 & 2 & 1 & -0.5 \\
			\hline 
			\hline  
			(20, 5) & 1.37 & -0.64 & 0.64 & 0.69 & 0.48 & -0.49 & 2.30 & 1.00 & -0.46 \\ 
			& (0.61) & (0.28) & (0.30) & (0.28) & (0.09) & (0.09) & (0.57) & (0.20) & (0.25) \\ 
			& (0.56) & (0.22) & (0.22) & (0.25) & (0.07) & (0.07) & (0.38) & (0.17) & (0.18) \\ 
			
			(20, 20) & 1.05 & -0.51 & 0.51 & 0.54 & 0.50 & -0.50 & 2.02 & 0.99 & -0.50 \\ 
			& (0.48)  & (0.09) & (0.09) & (0.23) & (0.03) & (0.03) & (0.38) & (0.18) & (0.19) \\ 
			& (0.46) & (0.08) & (0.08) & (0.23) & (0.03) & (0.03) & (0.33) & (0.16) &(0.16) \\ 
			
			(20, 50) & 0.99 & -0.50 & 0.51 & 0.52 & 0.50 & -0.50 & 1.97 & 0.98 & -0.48 \\ 
			& (0.45) & (0.05) & (0.05) & (0.23) & (0.02) & (0.02) & (0.35) & (0.16) & (0.19) \\ 
			& (0.44) & (0.05) & (0.05) & (0.22) & (0.02) & (0.02) & (0.32) & (0.16) & (0.17) \\ 
			
			(100, 5) & 1.27 & -0.59 & 0.58 & 0.64 & 0.49 & -0.49 & 2.12 & 1.03 & -0.51 \\ 
			& (0.25) & (0.11)  & (0.11) & (0.11) & (0.04) & (0.04) & (0.22) & (0.09) & (0.10) \\ 
			& (0.23) & (0.09) & (0.09) & (0.11) & (0.03) & (0.03) & (0.15) & (0.07) & (0.08) \\ 
			
			(100, 20) & 1.04 & -0.51 & 0.52 & 0.54 & 0.50 & -0.50 & 2.01 & 1.00 & -0.50 \\ 
			& (0.20) & (0.04) & (0.03) & (0.10) & (0.01) & (0.01) & (0.16) & (0.07) & (0.08) \\ 
			& (0.21) & (0.04) & (0.04) & (0.10) & (0.01) & (0.01) & (0.14) & (0.07) & (0.08) \\
			
			(100, 50) & 1.03 & -0.50 & 0.50 & 0.51 & 0.50 & -0.50 & 2.00 & 1.00 & -0.51 \\ 
			& (0.21) & (0.03) & (0.03) & (0.12) & (0.01) & (0.01) & (0.16) & (0.08) & (0.09) \\
			& (0.21) & (0.02) & (0.02) & (0.13) & (0.01) & (0.01) & (0.14) & (0.07) & (0.07) \\
			
			\hline
		\end{tabular}
	\end{center}
\end{table}	

\begin{table}[ht]
	\caption{Averaged coefficient estimates, standard deviations (SE) and the average standard errors (SEE) for the simulation scenario with dispersion parameters $\sigma_\beta = 0.5$, $\sigma_\alpha = 0.25$ and $\rho = -0.5$ for $25\%$ censoring rate.} 
	\footnotesize
	\begin{center}
		\begin{tabular}{||lccccccccc||} 
			\hline
			& $\hat{\beta}_0$ & $\hat{\beta}_1$ & $\hat{\beta}_2$ & $\hat{\alpha}_0$ & $\hat{\alpha}_1$ & $\hat{\alpha}_2$ & $\hat{\sigma_\beta}$ & $\hat{\sigma_\alpha}$ & $\hat{\rho}$\\	
			& Mean  & Mean  & Mean & Mean & Mean & Mean  & Mean  & Mean & Mean \\
			&  (SE) & (SE) & (SE) & (SE) & (SE) & (SE) & (SE) & (SE) & (SE) \\
			($q_i, n_i$)&  (SEE) & (SEE) & (SEE) & (SEE) & (SEE) & (SEE) & (SEE) & (SEE) & (SEE) \\
			True & 1 & -0.5 & 0.5 & 0.5 & 0.5 & -0.5 & 0.5 & 0.25 & -0.5 \\
			\hline 
			\hline
			(20, 5) & 1.22 & -0.58 & 0.57 & 0.62 & 0.49 & -0.50 & 0.62 & 0.28 & -0.35 \\ 
			& (0.26) & (0.23) & (0.22) & (0.13) & (0.10) & (0.10) & (0.25) & (0.13) & (0.60) \\ 
			& (0.22) & (0.19) & (0.19) & (0.12) & (0.09) & (0.09) & (0.12) & (0.06) & (0.20) \\ 
			
			(20, 20) & 1.07 & -0.52 & 0.52 & 0.53 & 0.50 & -0.50 & 0.50 & 0.24 & -0.50 \\ 
			& (0.13) & (0.08) & (0.08) & (0.08) & (0.04) & (0.04) & (0.13) & (0.07) & (0.35) \\ 
			& (0.14) & (0.08) & (0.08) & (0.07) & (0.04) & (0.04) & (0.09) & (0.04) & (0.17) \\
			
			(20, 50) & 1.02 & -0.50 & 0.51 & 0.51 & 0.50 & -0.50 & 0.49 & 0.25 & -0.51 \\ 
			& (0.12) & (0.05) & (0.05) & (0.06) & (0.03) & (0.03) & (0.09) & (0.05) & (0.24) \\ 
			& (0.12) & (0.05) & (0.05) & (0.06) & (0.03) & (0.03) & (0.08) & (0.04) & (0.17) \\
			
			(100, 5) & 1.16 & -0.54 & 0.54 & 0.56 & 0.50 & -0.49 & 0.57 & 0.26 & -0.43 \\ 
			& (0.10) & (0.08) & (0.08) & (0.06) & (0.04) & (0.04) & (0.10) & (0.07) & (0.35) \\ 
			& (0.09) & (0.08) & (0.08) & (0.05) & (0.04) & (0.04) & (0.05) & (0.02) & (0.11) \\ 
			
			(100, 20) & 1.04 & -0.51 & 0.51 & 0.52 & 0.50 & -0.50 & 0.50 & 0.25 & -0.50 \\ 
			& (0.06) & (0.04) & (0.04) & (0.03) & (0.02) & (0.02) & (0.05) & (0.03) & (0.15) \\ 
			& (0.06) & (0.03) & (0.03) & (0.03) & (0.02) & (0.02) & (0.04) & (0.02) & (0.08) \\ 	
			
			(100, 50) & 1.02 & -0.50 & 0.50 & 0.51 & 0.50 & -0.50 & 0.50 & 0.25 & -0.50 \\ 
			& (0.05) & (0.02) & (0.02) & (0.03) & (0.01) & (0.01) & (0.04) & (0.02) & (0.11) \\ 
			& (0.05) & (0.02) & (0.02) & (0.03) & (0.01) & (0.01) & (0.04) & (0.02) & (0.08) \\ 
			\hline
		\end{tabular}
	\end{center}
\end{table}

\begin{table}[ht]
	\caption{Averaged coefficient estimates, standard deviations (SE) and the average standard errors (SEE) for the simulation scenario with dispersion parameters $\sigma_\beta = 1$, $\sigma_\alpha = 0.5$ and $\rho = 0.5$ for $25\%$ censoring rate.}
	\footnotesize
	\begin{center}
		\begin{tabular}{||lccccccccc||} 
			\hline
			& $\hat{\beta}_0$ & $\hat{\beta}_1$ & $\hat{\beta}_2$ & $\hat{\alpha}_0$ & $\hat{\alpha}_1$ & $\hat{\alpha}_2$ & $\hat{\sigma_\beta}$ & $\hat{\sigma_\alpha}$ & $\hat{\rho}$\\	
			& Mean  & Mean  & Mean & Mean & Mean & Mean  & Mean  & Mean & Mean \\
			&  (SE) & (SE) & (SE) & (SE) & (SE) & (SE) & (SE) & (SE) & (SE) \\
			($q_i, n_i$)&  (SEE) & (SEE) & (SEE) & (SEE) & (SEE) & (SEE) & (SEE) & (SEE) & (SEE) \\
			True & 1 & -0.5 & 0.5 & 0.5 & 0.5 & -0.5 & 1 & 0.5 & 0.5 \\ 
			\hline
			\hline
			
			(20, 5) & 1.34 & -0.62 & 0.61 & 0.66 & 0.48 & -0.48 & 1.21 & 0.54 & 0.47 \\ 
			&  (0.39) &  (0.29) &  (0.28) &  (0.17) &  (0.12) &  (0.12) &  (0.38) &  (0.17) &  (0.37) \\ 
			&  (0.33) &  (0.21) &  (0.21) &  (0.16) &  (0.09) &  (0.09) &  (0.21) &  (0.10) &  (0.16) \\ 
			
			(20, 20) & 1.08 & -0.51 & 0.51 & 0.54 & 0.50 & -0.50 & 1.00 & 0.50 & 0.49 \\ 
			&  (0.25) &  (0.08) &  (0.08) &  (0.13) &  (0.04) &  (0.04) &  (0.19) &  (0.10) &  (0.21) \\ 
			&  (0.24) &  (0.08) &  (0.08) &  (0.12) &  (0.04) &  (0.04) &  (0.16) &  (0.08) &  (0.17) \\ 
			(20, 50) & 1.04 & -0.51 & 0.50 & 0.51 & 0.50 & -0.50 & 0.98 & 0.49 & 0.47 \\ 
			&  (0.23) &  (0.05) &  (0.05) &  (0.12) &  (0.02) &  (0.03) &  (0.17) &  (0.09) &  (0.19) \\ 
			&  (0.22) &  (0.05) &  (0.05) &  (0.11) &  (0.02) &  (0.02) &  (0.16) &  (0.08) &  (0.17) \\ 
			(100, 5) & 1.22 & -0.56 & 0.56 & 0.61 & 0.48 & -0.49 & 1.09 & 0.52 & 0.49 \\ 
			&  (0.15) &  (0.10) &  (0.10) &  (0.07) &  (0.04) &  (0.04) &  (0.13) &  (0.063) &  (0.13) \\ 
			&  (0.13) &  (0.09) &  (0.09) &  (0.07) &  (0.04) &  (0.04) &  (0.08) &  (0.04) &  (0.08) \\ 
			(100, 20) & 1.06 & -0.51 & 0.51 & 0.53 & 0.50 & -0.50 & 1.00 & 0.50 & 0.48 \\ 
			&  (0.11) &  (0.04) &  (0.04) &  (0.06) &  (0.02) &  (0.02) &  (0.08) &  (0.04) &  (0.09) \\ 
			&  (0.11) &  (0.04) &  (0.04) &  (0.05) &  (0.02) &  (0.02) &  (0.07) &  (0.04) &  (0.08) \\ 
			(100, 50) & 1.03 & -0.50 & 0.50 & 0.51 & 0.50 & -0.50 & 1.00 & 0.50 & 0.49 \\ 
			&  (0.10) &  (0.02) &  (0.02) &  (0.05) &  (0.01) &  (0.01) &  (0.07) &  (0.04) &  (0.08) \\ 
			&  (0.10) &  (0.02) &  (0.02) &  (0.05) &  (0.01) &  (0.01) &  (0.07) &  (0.04) &  (0.08) \\ 
			\hline
		\end{tabular}
	\end{center}
\end{table}

\begin{table}[ht]
	\caption{Averaged coefficient estimates, standard deviations (SE) and the average standard errors (SEE) for the simulation scenario with dispersion parameters $\sigma_\beta = 1$, $\sigma_\alpha = 0.5$, $\rho = -0.5$, $25\%$ censoring rate and varying cluster sizes ($30\%$ of the clusters are of size $5$, $40\%$ are of size $20$ and the remaining $30\%$ are of size $50$).}
	\footnotesize
	\begin{center}
		\begin{tabular}{||lccccccccc||}
			\hline
			& $\hat{\beta}_0$ & $\hat{\beta}_1$ & $\hat{\beta}_2$ & $\hat{\alpha}_0$ & $\hat{\alpha}_1$ & $\hat{\alpha}_2$ & $\hat{\sigma_\beta}$ & $\hat{\sigma_\alpha}$ & $\hat{\rho}$\\	
			& Mean  & Mean  & Mean & Mean & Mean & Mean  & Mean  & Mean & Mean \\
			&  (SE) & (SE) & (SE) & (SE) & (SE) & (SE) & (SE) & (SE) & (SE) \\
			($q_i$)&  (SEE) & (SEE) & (SEE) & (SEE) & (SEE) & (SEE) & (SEE) & (SEE) & (SEE) \\
			True & 1 & -0.5 & 0.5 & 0.5 & 0.5 & -0.5 & 1 & 0.5 & -0.5 \\ 
			\hline
			\hline
			(20) & 1.1 & -0.51 & 0.51 & 0.54 & 0.50 & -0.50 & 1.03 & 0.50 & -0.47 \\ 
			&  (0.26) &  (0.07) &  (0.07) &  (0.12) &  (0.04) &  (0.04) &  (0.20) &  (0.10) &  (0.24) \\ 
			&  (0.25) &  (0.07) &  (0.07) &  (0.12) &  (0.03) &  (0.03) &  (0.17) &  (0.08) &  (0.17) \\ 
			(100) & 1.08 & -0.51 & 0.51 & 0.52 & 0.50 & -0.50 & 1.00 & 0.50 & -0.50 \\ 
			&  (0.11) &  (0.03) &  (0.03) &  (0.06) &  (0.02) &  (0.01) &  (0.09) &  (0.05) &  (0.10) \\ 
			&  (0.11) &  (0.03) &  (0.03) &  (0.06) &  (0.01) &  (0.01) &  (0.07) &  (0.04) &  (0.08) \\ 
			\hline
		\end{tabular}
	\end{center}
\end{table}

\begin{table}[ht]
	\caption{Averaged coefficient estimates, standard deviations (SE) and the average standard errors (SEE) for the simulation scenario with dispersion parameters $\sigma_\beta = 1$, $\sigma_\alpha = 0.5$ and $\rho = -0.5$ for $50\%$ censoring rate.}
	\footnotesize
	\begin{center}
		\begin{tabular}{||lccccccccc||} 
			\hline
			& $\hat{\beta}_0$ & $\hat{\beta}_1$ & $\hat{\beta}_2$ & $\hat{\alpha}_0$ & $\hat{\alpha}_1$ & $\hat{\alpha}_2$ & $\hat{\sigma_\beta}$ & $\hat{\sigma_\alpha}$ & $\hat{\rho}$\\	
			& Mean  & Mean  & Mean & Mean & Mean & Mean  & Mean  & Mean & Mean \\
			&  (SE) & (SE) & (SE) & (SE) & (SE) & (SE) & (SE) & (SE) & (SE) \\
			($q_i, n_i$)&  (SEE) & (SEE) & (SEE) & (SEE) & (SEE) & (SEE) & (SEE) & (SEE) & (SEE) \\
			True & 1 & -0.5 & 0.5 & 0.5 & 0.5 & -0.5 & 1 & 0.5 & -0.5 \\ 
			\hline
			\hline
			(20, 5) & 1.59 & -0.79 & 0.79 & 0.68 & 0.45 & -0.44 & 1.55 & 0.61 & -0.04 \\ 
			& (0.72) & (0.66) & (0.65)  & (0.27) & (0.20)  & (0.21) & (0.66) & (0.27) & (0.60) \\ 
			& (0.49) & (0.39) & (0.38) & (0.20) & (0.13) & (0.13) & (0.29) & (0.12) & (0.19) \\ 
			
			(20, 20) & 1.11 & -0.52 & 0.51 & 0.53 & 0.50 & -0.50 & 0.99 & 0.50 & -0.48 \\ 
			& (0.29)  & (0.16) & (0.15) & (0.14) & (0.06) & (0.07) & (0.22) & (0.12) & (0.29) \\ 
			& (0.26) & (0.14) & (0.14) & (0.13) & (0.06) & (0.06) & (0.17) & (0.08) & (0.18) \\ 
			
			(20, 50) & 1.02 & -0.51 & 0.51 & 0.52 & 0.50 & -0.50 & 0.98 & 0.50 & -0.49 \\ 
			& (0.22) & (0.08) & (0.08) & (0.12) & (0.04) & (0.04) & (0.19) & (0.09) & (0.23) \\ 
			& (0.24) & (0.08) & (0.08) & (0.12) & (0.04) & (0.04) & (0.16) & (0.08) & (0.17) \\ 
			
			(100, 5) & 1.38 & -0.62 & 0.63 & 0.63 & 0.47 & -0.47 & 1.16 & 0.56 & -0.24 \\ 
			& (0.22) & (0.23) & (0.22) & (0.11) & (0.08) & (0.08) & (0.22) & (0.11) & (0.28) \\ 
			& (0.18) & (0.16) & (0.16) & (0.09) & (0.06) & (0.06) & (0.09) & (0.05) & (0.12) \\ 
			
			(100, 20) & 1.08 & -0.52 & 0.53 & 0.53 & 0.49 & -0.49 & 0.98 & 0.51 & -0.48 \\ 
			& (0.12) & (0.07) & (0.07) & (0.06) & (0.03) & (0.03) & (0.10) & (0.05) & (0.13) \\ 
			& (0.11) & (0.06) & (0.06) & (0.06) & (0.03) & (0.03) & (0.07) & (0.04) & (0.08) \\ 
			
			(100, 50) & 1.03 & -0.51 & 0.51 & 0.52 & 0.50 & -0.50 & 0.99 & 0.50 & -0.51 \\ 
			& (0.11) & (0.04) & (0.04) & (0.06) & (0.02) & (0.02) & (0.09) & (0.04) & (0.09) \\ 
			& (0.11) & (0.04) & (0.04) & (0.05) & (0.02) & (0.02) & (0.07) & (0.04) & (0.08) \\ 
			\hline
		\end{tabular}
	\end{center}
\end{table}

\end{document}